# ON THE INTEGRALS INVARIANT WITH RESPECT TO A PARAMETER

V.I. Fabrikant,
prisoner #167 932 D, Archambault Jail,
242 Montee Gagnon
Ste-Anne-Des-Plaines, Quebec Canada J0N1H0

**Abstract.** Author presents a study of certain category of the integrals, which might look quite difficult to compute, but in fact are easily computable, because they do not depend on the parameter in the integrand. As simple and elementary the procedure is, the author discovered that the best of symbolic manipulation packages presently available fail quite miserably, when asked to compute these integrals. The author suggests proper modifications in the algorithms used for symbolic integration.

## INTRODUCTION

The idea of the article appeared when the author came across the integral 4.538.1 from Gradshteyn and Ryzhik (1994), which reads:

$$\int_0^\infty \operatorname{arctg}(x^2)\frac{dx}{1+x^2} = \int_0^\infty \operatorname{arctg}(x^3)\frac{dx}{1+x^2} = \int_0^\infty \operatorname{arcctg}(x^2)\frac{dx}{1+x^2} =$$

$$= \int_0^\infty \operatorname{arcctg}(x^3)\frac{dx}{1+x^2} = \frac{\pi^2}{8}. \tag{1}$$

The integral looked peculiar, so it was natural to try different powers of the argument in arctg, and the result was invariably the same, even when the power was a non-integer. It was natural then to presume that

$$\int_0^\infty \operatorname{arctg}(x^a)\frac{dx}{1+x^2} = \frac{\pi^2}{8}, \tag{2}$$

where *a* is an arbitrary real number. We have here the case, where the integrand depends on parameter *a* and the result of integration is invariant with respect to this parameter.

The investigation of this kind of integrals is presented in the next section.

## STUDY

First, we present 2 different proofs of correctness of (2). We split the interval of integration in 2: from 0 to 1 and from 1 to ∞; in the second integral thus obtained, we make formal substitution of *x* by 1/*x*. The result is

$$\int_0^\infty \operatorname{arctg}(x^a)\frac{dx}{1+x^2} = \int_0^1 \operatorname{arctg}(x^a)\frac{dx}{1+x^2} + \int_1^\infty \operatorname{arctg}(x^a)\frac{dx}{1+x^2} =$$

$$= \int_0^1 \operatorname{arctg}(x^a)\frac{dx}{1+x^2} + \int_0^1 \operatorname{arctg}(1/x^a)\frac{dx}{1+x^2}. \tag{3}$$

Now we can use the identity $\arctg(z)+\arctg(1/z) = \pi/2$, and correctness of (2) becomes obvious.

And here is yet another manner to prove correctness of (2). We denote

$$\int_0^\infty \arctg(x^a)\frac{dx}{1+x^2} = f(a). \tag{4}$$

Presume that $f(a)$ can be expanded in Taylor series. We compute

$$f(0) = \frac{\pi}{4}\int_0^\infty \frac{dx}{1+x^2} = \frac{\pi^2}{8}. \tag{5}$$

Now we need to prove that all the derivatives of $f(a)$ with respect to $a$ are equal to zero. Here are the first four derivatives:

$$\frac{df(a)}{da} = \int_0^\infty \frac{x^a \ln x}{1+x^{2a}}\frac{dx}{1+x^2}. \tag{6}$$

$$\frac{d^2 f(a)}{da^2} = \int_0^\infty \frac{x^a(1-x^{2a})\ln^2 x}{(1+x^{2a})^2}\frac{dx}{1+x^2}. \tag{7}$$

$$\frac{d^3 f(a)}{d^3 a} = \int_0^\infty \frac{x^a(1-6x^{2a}+x^{4a})\ln^3 x}{(1+x^{2a})^3}\frac{dx}{1+x^2}. \tag{8}$$

$$\frac{d^4 f(a)}{d^4 a} = \int_0^\infty \frac{x^a(1-23x^{2a}+23x^{4a}-x^{6a})\ln^4 x}{(1+x^{2a})^4}\frac{dx}{1+x^2}. \tag{9}$$

From (6-9) we may conclude that the odd derivatives are zero, because the odd power of $\ln(x)$ is negative for $x<1$ and positive for $x>1$; the derivatives of even order are zero, because the formal substitution of $x$ by $1/x$ yields the same expression, but with opposite sign. We now need to prove that this will be true for a derivative of arbitrary order. We achieve it by the method of mathematical induction. We presume that the $n$-th derivative has the form:

$$\frac{d^n f(a)}{d^n a} = \int_0^\infty \frac{\ln^n x}{(1+x^{2a})^n}\sum_{k=1}^n C_k x^{(2k-1)a} \frac{dx}{1+x^2}, \tag{10}$$

with $C_m = C_{n-m+1}$ for odd $n$ and $C_m = -C_{n-m+1}$ for even $n$. If this is true for $n$, then we need to prove that the same would be true for $n+1$. We write $n+1$ derivative as

$$\frac{d^{n+1} f(a)}{d^{n+1} a} = \int_0^\infty \frac{\ln^{n+1} x}{(1+x^{2a})^{n+1}}\sum_{k=1}^{n+1} D_k x^{(2k-1)a} \frac{dx}{1+x^2}, \tag{11}$$

The differentiation of (10) gives us the following relationship between $C_k$ and $D_k$:

$$D_1 = C_1, \qquad D_{n+1} = -C_n, \qquad D_m = (2m-1)C_m - (2n-2m+3)C_{m-1},$$

$$D_{n-m+2} = (2n-2m+3)C_{n-m+2} - (2m-1)C_{n-m+1}. \tag{12}$$

From (12) we may conclude that the first coefficient of the polynomial in the $n+1$ derivative remains unchanged, which means it is equal to 1; the last term of the polynomial is equal and opposite to the last term of the previous derivative, which means that if it was 1, then it becomes -1 and vice-versa. If $C_m = C_{n-m+1}$, then from (12) it follows that $D_m = -D_{n-m+2}$ and vice-versa, as it was postulated. The proof is completed. We obtained an infinite set of integrands, such that the result of integration in the interval from 0 to  is always zero.

The category of integrals invariant with respect to a parameter is not limited to polynomials similar to (10). Here is an example of the integral, which appeared naturally in the research of the author:

$$\int_0^{2\pi} \ln \frac{z^2 + (a + r\cos\varphi)^2}{z^2 + (a - r\cos\varphi)^2} d\varphi. \tag{13}$$

At first, it looks like the result of integration should be a function of 3 parameters $z$, $a$, and $r$, while in fact, it is equal zero. This can be easily shown by splitting the interval of integration in two: from 0 to  and from  to 2 ; in the second integral we make a formal substitution of  by  + .

## DISCUSSION

The analysis presented above is rather simple and elementary, and the author was very surprised to discover that it seems to have escaped attention of other scientists. This becomes evident from examination of existing tables of integrals, as well as by trials of existing software packages capable of symbolic integration. Here are our findings.

The following integral was found in Gradshteyn and Ryzhik (1994, formula 4.127.7)

$$\int_0^\infty \frac{(\ln x)^{2n+1}}{1 + bx + x^2} dx = 0. \tag{14}$$

The author did not seem to realize that much more general integrand can be presented, namely,

$$\int_0^\infty \frac{(\ln x)^{2n+1}}{1 + bx + x^2} \frac{\sum_{k=0}^{2m} p_k x^k}{\sum_{k=0}^{2m} q_k x^k} dx = 0, \text{ with } n \text{ and } m \text{ arbitrary integers}, \quad p_k = p_{2m-k} \text{ and } q_k = q_{2m-k}. \tag{15}$$

Yet another variation of similar integral can be written as

$$\int_0^\infty \frac{(\ln x)^{2n}}{1+bx+x^2} \frac{\sum_{k=0}^{2m+1} p_k x^k}{\sum_{k=0}^{2m+1} q_k x^k} dx = 0, \quad \text{with } n \text{ and } m \text{ arbitrary integers, } p_k = -p_{2m+1-k} \text{ and } q_k = -q_{2m+1-k}. \tag{16}$$

We found yet another invariant integral in Gradshteyn and Ryzhik (1994, formula 4.297.2)

$$\int_0^\infty \ln\left(\frac{ax+b}{bx+a}\right) \frac{dx}{(1+x)^2} = 0. \tag{17}$$

The authors of the book did not seem to realize that a more general result could be claimed, namely,

$$\int_0^\infty \ln\left(\frac{\sum_{k=0}^m p_k x^k}{\sum_{k=0}^m p_{m-k} x^k}\right) \frac{dx}{1+bx+x^2} = 0, \quad \text{with } m \text{ arbitrary integer.} \tag{18}$$

Now we turn to the software packages capable of performing symbolic integration. Mathematica was unable to integrate (2); it did compute (6), but failed to integrate any of higher derivatives (7-9). It also failed to integrate (13). Maple failed in all computations.

It is obvious that the designers of the symbolic software missed some important steps in the algorithm of symbolic integration. We suggest the following steps to be incorporated in the algorithm.

1. Numerical computation of the integral for 2 different sets of parameters to see whether it is invariant.
2. If yes, then split the interval of integration in two and make a formal change of variables in the second integral to make the intervals of integration identical. In the case the interval of integration is from 0 to  the subintervals should be from 0 to 1 and from 1 to  ; in the case of interval of integration from zero to 2 , the subintervals should be from 0 to  and from  to 2 , etc. Simplify combined integrand, having used the identities $\ln(x) = -\ln(1/x)$ or $\text{arctg}(x) + \text{arctg}(1/x) = /2$.

**Conclusion.** Simple and elementary analysis presented here seems to have escaped attention of other scientists and implementation of its results will improve both quality of tables of integrals and algorithms of the software designed to perform symbolic integration.